# Functional and structural ophthalmic imaging using noncontact multimodal photoacoustic remote sensing microscopy and optical coherence tomography

*Zohreh Hosseinaee [1], Nima Abbasi F. [1], Nicholas Pellegrino[1], Layla Khalili[1], Lyazzat Mukhangaliyeva[1], Parsin Haji Reza [1]\**

1. PhotoMedicine Labs, Department of System Design Engineering, University of Waterloo, 200 University Ave W, Waterloo, ON, N2L 3G1, Canada

*Email: phajireza@uwaterloo.ca

**Abstract**

Early diagnosis of ocular diseases improves the understanding of pathophysiology and aids in accurate monitoring and effective treatment. Advanced, multimodal ocular imaging platforms play a crucial role in visualization of ocular components and provide clinicians with a valuable tool for evaluating various eye diseases. Here, for the first time we present a non-contact, multiwavelength photoacoustic remote sensing (PARS) microscopy and swept-source optical coherence tomography (SS-OCT) for *in-vivo* functional and structural imaging of the eye. The system provides complementary imaging contrasts of optical absorption and optical scattering, and is used for simultaneous, non-contact, *in-vivo* imaging of murine eye. Results of vasculature and structural imaging as well as melanin content in the retinal pigment epithelium layer are presented. Multiwavelength PARS microscopy using Stimulated Raman Scattering is applied to enable *in-vivo*, non-contact oxygen saturation estimation in the ocular tissue. The reported work may be a major step towards clinical translation of ophthalmic technologies and has the potential to advance the diagnosis and treatment of ocular diseases.

**1       Introduction**

The ability to precisely detect and measure chromophore concentration in ocular tissue is crucial for improving investigations and diagnoses of major eye diseases. Several studies have demonstrated the vital role of ocular oxygen saturation ($SO_2$) and melanin concentration in various prevalent eye diseases[1]. For example, retinal oxygen saturation has been shown to be abnormal in diabetic retinopathy (DR)[2,3], glaucoma diseases[4,5], and retinal vein occlusions[6,7]. In addition, melanin loss in the retinal pigment epithelium (RPE) layer is shown to be a major factor affecting the progression of age-related macular degeneration (AMD)[8,9]. Therefore, accurate measurement of chromophore concentrations in the ocular environment could potentially enable improved measurement of biomarkers for early diagnosis of eye diseases[10]. In the past decades, extensive efforts have been made to accurately measure $SO_2$ and melanin concentration in the ocular environment. The gold standard for measuring retinal $SO_2$ in ophthalmic research uses intra-retinal electrodes to directly measure the oxygen tension[11], however, the invasive nature of the method limits its clinical applications, thus it can only be used in human subjects during surgical procedures[12]. Additionally, it is nearly impossible to map the amount of oxygenation over a large area using this method. Imaging-based methods are required in order to visualize the spatial distribution of oxygen in ocular blood vessels. Phosphorescence lifetime imaging has been used to map oxygenation in the mouse eye. Unfortunately, the need to introduce fluorescent probes into systematic circulation makes the method inappropriate for human practice[13]. Magnetic resonance imaging (MRI) can also detect retinal oxygen distribution non-invasively in humans, however, it has low resolution and offers only qualitative measurement[14]. Recently, researchers are focused on optical imaging-based methods to extract blood oxygen saturation. Optical measurement of $SO_2$ is possible because the two forms of hemoglobin, oxy- and deoxyhemoglobin ($HbO_2$ and Hb), have distinct optical absorption properties. The differences in the absorption spectra of oxy- and deoxyhemoglobin can be utilized to quantify the relative amounts of $HbO_2$ and Hb in the blood[15]. Several optical imaging methods, including multiwavelength fundus imaging, scanning laser ophthalmoscopy and visible light optical coherence tomography (OCT), have all been investigated to measure ocular oxygen saturation[12]. These methods, however, rely on measuring the backscattered photons from blood vessels to quantify the absorption of specific chromophores inside the tissue[16,17]. Therefore, they are sensitive to local geometrical parameters, such as retinal thickness, vessel diameters, and retinal pigmentation and may result in biased estimations[1].

Among various optical imaging modalities, photoacoustic microscopy (PAM) offers unique imaging contrast of optical absorption. In other words, any imaging target that absorbs light energy, can be imaged using PAM. Biological tissues have endogenous chromophores that can be exploited as imaging targets. For example, the absorption peak of DNA/RNA is in the ultraviolet spectral region, and hemoglobin and melanin mainly absorb light in the visible and near-infrared (NIR) spectral ranges[18]. This unique imaging ability makes PAM a favorable candidate for various functional and molecular imaging applications and measuring chromophore concentration[19]. Over the past decades, photoacoustic ophthalmoscopy has been applied for visualizing hemoglobin and melanin content in ocular tissue[20], quantifying ocular $SO_2$[21], and measuring the metabolic rate of oxygen consumption ($MRO_2$)[22]. Despite all these advantages offered by PAM devices, a major limitation arises from their need to be in contact with the ocular tissue[18]. This physical contact may increase the risk of infection and may cause patient discomfort. Furthermore, this contact-based imaging approach applies pressure to the eye and introduces barriers to oxygen diffusion. Thus, it has a crucial influence on the physiological and pathophysiological balance of ocular vasculature function, and it is not capable of studying dynamic processes under close conditions to normality[23].

In 2017 Haji Reza et al. developed photoacoustic remote sensing (PARS) microscopy for non-contact, non-interferometric detection of photoacoustic signals[24]. PARS microscopy can be considered as the non-contact, all-optical version of optical resolution PAM (OR-PAM), where the acoustically coupled ultrasound transducer is replaced with a co-focused probe beam. This all-optical detection scheme allows the system to measure the photoacoustic pressure waves at the subsurface origin where the pressure is maximum. Besides optical absorption imaging contrast, PARS also offers optical scattering contrast through its probe beam and can be considered as a dual-contrast imaging modality. Using only the probe beam, the PARS microscope can act as a confocal microscope to visualize scattering information of the tissue. In functional studies such as $SO_2$ measurement an additional advantage of PARS microscopy over other optical imaging modalities comes from its sensitivity to both optical absorption and optical scattering imaging contrasts. In other words, the scattering information provided through the probe beam of PARS microscopy can be used the same way as other scattering-based imaging modalities such as fundus photography or OCT to measure the amount of absorption inside the tissue. In addition, the wavelength of PARS excitation beam can be tuned to target a specific chromophore inside the tissue. The technology has proved its potential over a short period of time in various biomedical applications, such as label-free histological imaging[25,26], SO$_2$ mapping and angiogenesis imaging[27]. Very recently, our group (PhotoMedicine Labs) demonstrated the first, non-contact, *in-vivo* photoacoustic imaging of ocular tissue using PARS microscopy[28].

In ophthalmic imaging applications, optical coherence tomography is a state-of-the-art imaging technique extensively used in preclinical and clinical applications for imaging both anterior and posterior parts of the eye[29]. Unlike photoacoustic imaging, OCT obtains its imaging contrast from optical scattering of internal tissue microstructures. Due to its interferometric nature, OCT provides depth-resolved scattering information and can be considered as an ideal companion for PARS microscopy for ophthalmic imaging applications. This combined multimodal imaging technology has the potential to provide chromophore selective absorption contrast in concert with depth-resolved scattering contrast in the ocular enviroment[30]. Recently, Martell et al.[31] reported a dual-modal PARS microscopy combined with spectral-domain OCT (SD-OCT) and applied it for *in-vivo* imaging of ear tissue.

To allow for *in-vivo,* non-contact, functional and structural ophthalmic imaging, here we have combined a multiwavelength PARS microscope with a swept source OCT system (SS-OCT). SS-OCT has been used in this study, as it provides extended imaging range, reduced sensitivity roll-off and improved light detection efficiency compared to SD-OCT counterparts. To the best of our knowledge, this is the first time that a swept-source OCT system is combined with an OR-PAM system in general (both contact-based OR-PAM and non-contact OR-PAM – i.e., PARS). In addition, for the first time we present dual-contrast PARS microscopy, where multiwavelength excitation is used for targeting absorption contrast and the probe beam is used for targeting scattering imaging contrast. Additionally, here by capitalizing on the distinct differences in absorption spectra of oxy- and deoxyhemoglobin, oxygen saturation is estimated in the ocular tissue. To our knowledge, this is the first time a non-contact photoacoustic system is used for *in-vivo* SO$_2$ measurement in the ocular environment. This reported work has the potential to advance the diagnosis and treatment of major eye diseases.

## 2    Methods:

### 2.1    *Stimulated Raman Scattering*

Previous multiwavelength OR-PAM studies commonly used dye lasers or optical parametric oscillators to obtain multiple wavelengths required for oxygen saturation measurements[32,33]. However, these sources suffer from low pulse repetition rates (PRR) of 10 Hz – 10 kHz and thus are not suitable for *in-vivo* ocular imaging applications. In this study to achieve multiwavelength PARS microscopy and provide a reasonable speed for eye imaging, the SRS effect is employed. SRS offers an effective approach for implementing high-speed and multiwavelength light sources[34]. Distinct SRS peaks are generated from inelastic nonlinear interactions between incoming photons and the molecules of the fiber itself[35]. Here, the output of the excitation laser was coupled into a 3.5-m polarization-maintaining single-mode fiber (PM-460 HP) using a fiber launch system. The coupling efficiency was ~65%, and four distinct optical wavelengths were observed at the output of the fiber. To evaluate the performance of the multiwavelength light source, a comprehensive study was conducted on the peak power values, damage threshold, and the temporal stability of the procedure[36]. Input power levels are limited by the fiber damage threshold. In our experiments, the input power was varied between 1 mW- 800 mW. The fiber damage was first observed at 100-kHz pulse-repetition-rate and 800 mW average input power. SRS efficiency could be affected by temperature and airflow changes, and random errors and drift in pulse energy[37]. To improve the stability of the SRS peaks, the fiber was kept in a temperature-controlled unit to isolate the airflow and maintain the temperature within ± 0.1°C. Additionally, at the output port of the fiber, which is connected to the collimator, the back-reflected light from the filter may also cause damage at the end of the fiber; this can be solved by introducing a small angle to the filter[34]. A fiber optic spectrometer measured the SRS peaks and confirmed the filtered wavelengths. Table 1 shows maximum powers measured for each wavelength at 100 kHz pulse-repetition-rate.

**Table 1:** Measured power of SRS peaks generated in 3.5m fiber and at 100kHz PRR.

| 3.5-m fiber (100 kHz) | | | | |
|---|---|---|---|---|
| **Generated Wavelength (nm)** | 532 | 545 | 558 | 573 |
| **Output Energy (nJ)** | 250 | 230 | 230 | 180 |

## *2.2 System Architecture*

Figure 1a demonstrates the experimental setup of the multimodal PARS-OCT system. Details of the PARS subsystem was explained in the previous study published by our group for non-contact ophthalmic imaging[28]. Briefly, a 532-nm 1.5 ns pulse-width, ytterbium-doped fiber laser (IPG Photonics) is coupled to a single mode optical fiber to generate SRS peaks. The collimated excitation beam passes through bandpass filters to select the desired wavelength and is directed toward the sample. The detection arm uses an 830-nm Superluminescent Diode with 20 nm full width at half maximum linewidth (SLD830S-A20, Thorlabs). A polarized beam splitter is used to transmit the majority of the forward light onto a quarter wave-plate, which transforms the linearly polarized light into circularly polarized light. The detection and excitation beams are then combined using a dichroic mirror. The co-aligned beams are directed toward a large-beam galvanometer scanning mirror system (GVS012/M, Thorlabs, Inc.). The beams are then directed to a telecentric pair set that provides uniform image intensity and improves the effective field-of-view. The excitation and detection beams are co-focused into the sample using a refractive objective lens (depending on the application the numerical aperture (NA) of the objective lens is switched between 0.26 and 0.4). The back-reflected light from the sample is collected via the same objective lens and guided towards the detection path. The quarter wave-plate transforms the reflected circularly polarized light back to linearly polarized light, which enables the polarized beam splitter to direct the back-reflected light towards the photodiode. A long-pass filter is used to block any residual 532 nm light. The 830 nm beam is then focused with an aspherical lens onto a balanced photodiode. The photodiode outputs are connected to a high-speed digitizer (CSE1442, Gage Applied, Lockport, IL, USA) that performs analog to digital signal conversion.

In the SS-OCT subsystem, a vertical cavity surface emitting laser (VCSEL) (Thorlabs, Inc.) is used as the light source. The laser is centered at ~1060 nm with 100 nm spectral bandwidth and frequency swept at 60 kHz which enables higher sensitivity compared to higher speed SS-OCT counterparts. A-line trigger, i. e. sweep trigger, was supplied by the light source and a K-linear sampling clock was provided by the Mach-Zehnder interferometer-based clock module integrated within the laser[38]. The output of the laser was connected to a custom fiber optic interferometer consisting of a circulator and a 50:50 fiber coupler that splits the light into the reference and sample arms. The reference arm consists of BK7 dispersion compensating prisms in the optical path and a translating mirror to set the zero delay. In the sample arm the collimated light is combined with the PARS excitation and probe beams and together are directed to the sample. The returning light from the sample and reference arms were interfered at the coupler and detected by the built-in dual balanced photodetector. The OCT signal was digitized by a high-speed A/D card (ATS9351, Alazar Technologies Inc., Pointe-Claire, QC, Canada). The raw OCT data was transmitted to a host computer through a PCI-Express interface. OCT system control was implemented in MATLAB platform to automatically control all the operations including system calibration, galvo-scanning, system synchronization, real-time imaging preview and data acquisition. The multimodal PARS-OCT system has two independent imaging modes. It can either performs as two standalone subsystems, or as a multimodal imaging unit to acquire simultaneous images. Figure 1b demonstrates the timing chart of the multimodal imaging system. Briefly, in the simultaneous imaging mode, at each A-line, and for each rising edge of sweep trigger signal, the OCT digitizer collects data during the sampling clock of the K-clock, and it stops collecting data during the dummy clock (frequency ~ 250 MHz). The sweep-trigger signal also generates an auxiliary signal, and the falling edge of this auxiliary signal is used as the external trigger for PARS excitation and data acquisition. As a result, during the time that OCT system is collecting data, the PARS excitation is off and it turns on during the backward sweep of the swept source laser, where the OCT digitizer has stopped collecting signal. This will ensure that the photoacoustic pressure waves induced by PARS excitation, will not affect the quality of the OCT signal.

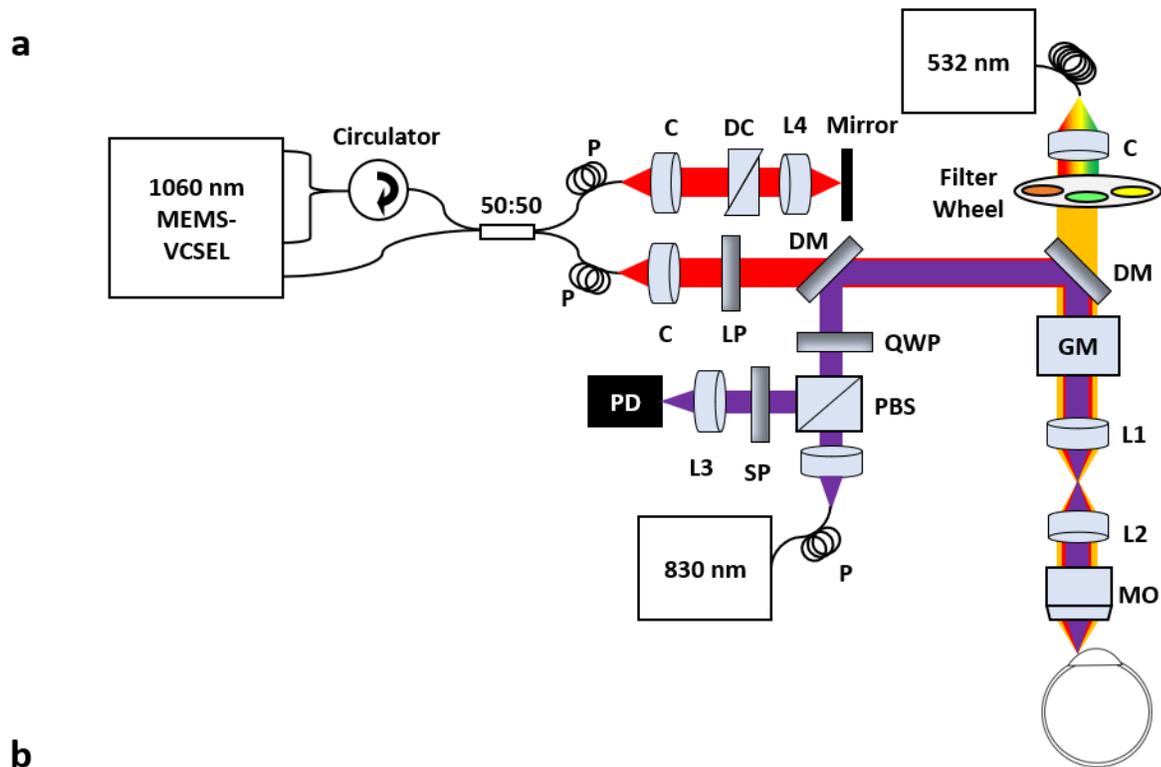

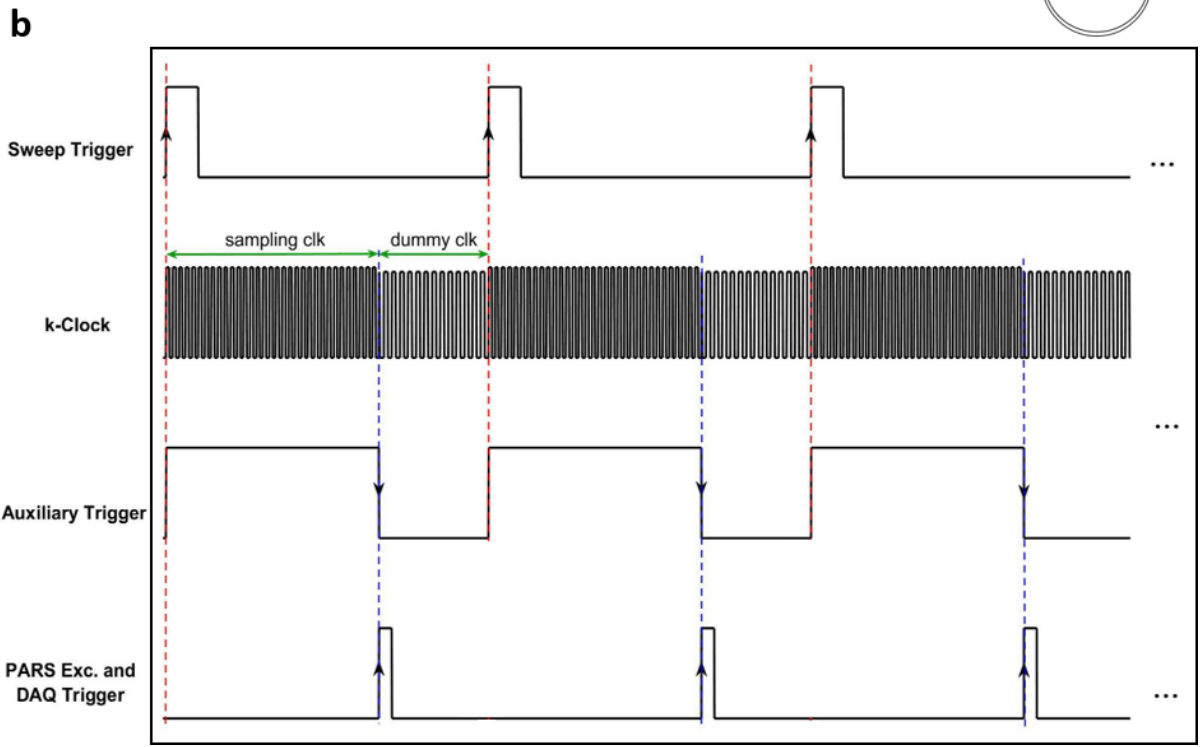

**Figure 1:** Simplified schematic and timing chart of the multimodal PARS-OCT system. Simplified system schematic. DM: Dichroic mirror, QWP: Quarter wave plate, PBS: Polarized beamsplitter, LP: Long pass filter, GM: Galvanometer mirrors, MO: microscope objective, L: Lens, C: Collimator, PD: Photodiode. DC: Dispersion compensation, P: Polarization controller (a). Multimodal imaging mode timing chart (b).

## 2.3  *Image Reconstruction*

All the PARS images shown in this manuscript were formed using a maximum amplitude projection (MAP) of each A-scan for each pixel of the *en-face* image. The images were produced by direct plotting from interpolated raw data using a Delaunay triangulation interpolation algorithm[39]. All images and signal processing steps were performed in the MATLAB environment. Scale bars in the field-of-view were calibrated using a 1951 USAF resolution test target.

For each OCT data set, 500 A-lines were acquired for each B-scan and on each cross-sectional position the slow scan axis traversed through 500 lateral positions. For each A-line trigger, 2448 sampling points were acquired to cover the resultant spectral interferogram, providing a depth ranging distance of ~12 mm. As a pre-processing step to extract complex data, the OCT reference spectrum was subtracted from the interference signal to remove DC bias, then Fourier transform was performed to extract the depth-resolved OCT signal. The top half of the Fourier transformed data was considered as valid complex data for further processing. Images were generated from the raw OCT data and numerically dispersion compensated up to the 5th order with a custom MATLAB algorithm[40]. No additional image post-processing was used for the OCT images presented in this paper. The volumetric and *en-face* images were generated from the 3D data sets with ImageJ[41].

## 2.4  *Animal Preparation*

All of the experimental procedures were carried out in conformity with the laboratory animal protocol approved by the Research Ethics Committee at the University of Waterloo and adhered to the ARVO statement for use of animals in ophthalmic and vision research. All sections of this report adhere to the ARRIVE Guidelines for reporting animal research. Nude mice and albino mice (NU/NU, Charles River, MA, USA) were imaged to demonstrate the *in-vivo* capabilities of the system. A custom-made animal holder was used to restrain the animal. The base of the animal holder was lined with a thermal pad in order to keep the animal body temperature between 36° and 38°C. Artificial tears were used frequently (~ every 5 minutes) to keep the cornea hydrated. Vital signs, such as respiration rates, heart rates and body temperature were monitored during the experiment.

*2.5  Ocular Light Safety*

Light safety is an important factor to consider in ocular imaging applications. In this study, the OCT light power on the cornea was measured to be ~1.5 mW centered at 1060 nm which is well within the ANSI safety limits for imaging human eye[42]. For the PARS system the detection power was ~ 2-3 mW which is in the range allowed by ANSI limits for imaging human eye. The PARS excitation pulse energy was in the range of 50-100 nJ. The ANSI standard was used to calculate the maximum permissible exposure (MPE) for ocular imaging [42,43]. Based on the ANSI standard, the MPE for a single laser pulse can be measured as $MPE_{SP}=5.0\ CE \times 10^{-7}=1.33\times 10^{-4}$ J/cm$^2$, where $C_E$ is a correction factor calculated as 267 according to the NA of the system. The repetitive pulse limit was then calculated as $MPE_{RP} = (n_{total})^{-0.25} \times MPE_{SP} = (5\times 10^5)^{-0.25} \times 1.33\times 10^{-4}$ J/cm$^2 = 5.12\times 10^{-6}$ J/cm$^2$, where $n_{total}$ is the total number of pulses during imaging (500,000 pixels in the image). The maximum permissible single laser pulse energy in a typical human pupil of 7 mm[44] was then calculated as $MPE_{RP} \times$ pupil area=1.93 µJ, which is 19 times higher than the pulse energy used in this experiment.

**3  Results and Discussion**

The imaging performance of the multimodal PARS-OCT system was evaluated by measuring its spatial resolution and signal-to-noise ratio (SNR). These results are presented in Figure 2. As mentioned earlier, PARS microscopy can be considered as a dual contrast imaging modality, providing both absorption and scattering imaging contrasts. The resolutions of both contrast mechanisms of PARS microscopy system were characterized by imaging 0.97 µm diameter microbead solution. The lateral resolution provided by PARS absorption contrast with a 0.4 NA objective lens was measured to be ~ 1.5 ± 0.1 µm, as shown in

Figure 2a. The black line shows the raw data, and the red line represents the first order Gaussian curve fit using the MATLAB curve fitting toolbox. Similarly, the resolution of PARS scattering contrast was measured to be ~ 3.1 ± 0.1 µm (Figure 2b).

The OCT axial resolution was measured experimentally by using a mirror as the test sample. The result is shown in Figure 2c. The red line point spread function (PSF) was measured after the coarse hardware dispersion compensation unit in the reference arm. The black line PSF in Figure 2c was measured after numerical dispersion compensation up to the 5$^{th}$ order using a custom MATLAB based algorithm. The full width half maximum of the axial PSF was 10.1 µm in free space, which corresponds to 7.3 µm in biological tissue, assuming an average refractive index of n = 1.38 and ignoring wavelength dependent local variation of the refractive index[45]. The SNR roll-off of the SS-OCT system was characterized by imaging a mirror in the sample arm at different positions. A neutral density filter with an optical density (OD) of 2.0 was used to reduce the signal intensity. The SNR in dB was calculated as the ratio of the A-scan peak height to the standard deviation of the noise floor. The maximum SNR of 100 dB was measured at ~ 100 µm away from the zero-delay line with incident power of ~ 1.5 mW. The SNR roll-off in free space was measured to be ~ 1dB over a scanning range of 1.3 mm (Figure 2d).

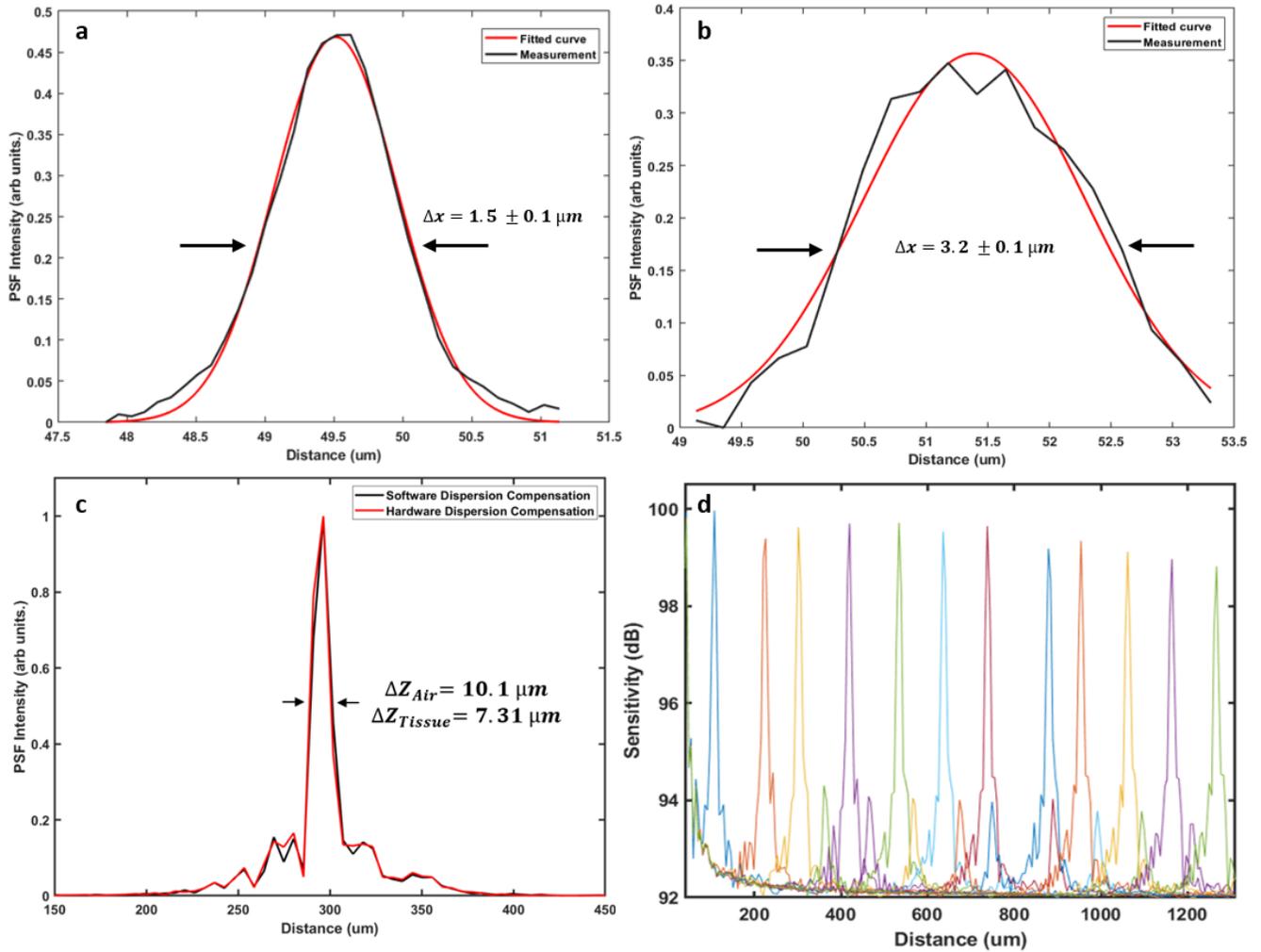

**Figure 2:** Imaging performance of the multimodal PARS-OCT system. PSF of PARS absorption contrast mechanism characterized by imaging 0.97 µm diameter microbead solution (a). PSF of PARS scattering contrast (b). Axial PSF of SS-OCT measured in free space (c). Sensitivity roll-off of SS-OCT (d)

To evaluate the *in-vivo* capabilities of our multimodal imaging system, we chose to first image mouse ear, a long-established animal model of cutaneous microcirculation that closely resembles that of human skin[46]. Since the multimodal system is configured similarly to a conventional optical microscope, it can be readily applied for studying other popular animal models such as zebra fish[47]. Figure 3b demonstrates a volumetric rendering of nude mouse ear covering ~ 2.5 mm × 2.5 mm area obtained *in-vivo* by SS-OCT system through a 0.26 NA microscope objective. The SS-OCT system provided a microanatomy of the avascular structures in the ear tissue (Figure 3c). From the OCT cross-sectional images we were able to estimate that the imaged region had a thickness between 200 to 300 µm, which agreed well with previous reports[48]. The orthogonal view of the skin (Figure 3d) clearly shows the structures of the ear due to the system's high resolution. The ear tissue consists of two skin layers separated by a layer of non-scattering cartilage, whereas, the epidermis, the outmost layer of the skin, tends to be more scattering. The junction between epidermis and dermis is clearly visible followed by the dermis where hair follicles, follicle shafts, follicle bulbs, and small blood and lymph vessels are located. The low-scattering regions embedded in the skin layers are most likely sebaceous glands. However, blood vessels are not evident on the OCT images. In contrast, the PARS absorption mechanism is good at locating vasculature with great details in the ear (Figure 3 e-g). For the mouse ear images the SNR, defined as the average of the maximum amplitude projection pixels in a region of interest over the standard deviation of the noise, was quantified as $36 \pm 3$ dB. The *in-vivo* resolution of the mouse ear images were measured as approximately $2.4 \pm 0.5$ µm. The *in-vivo* performance of the PARS scattering mechanism is presented in Figure 3h. Due to its shorter central wavelength PARS scattering mechanism offers higher lateral resolution compared to the OCT system, and microscopic epidermal ridges are evident in the MAP image acquired using the PARS system.

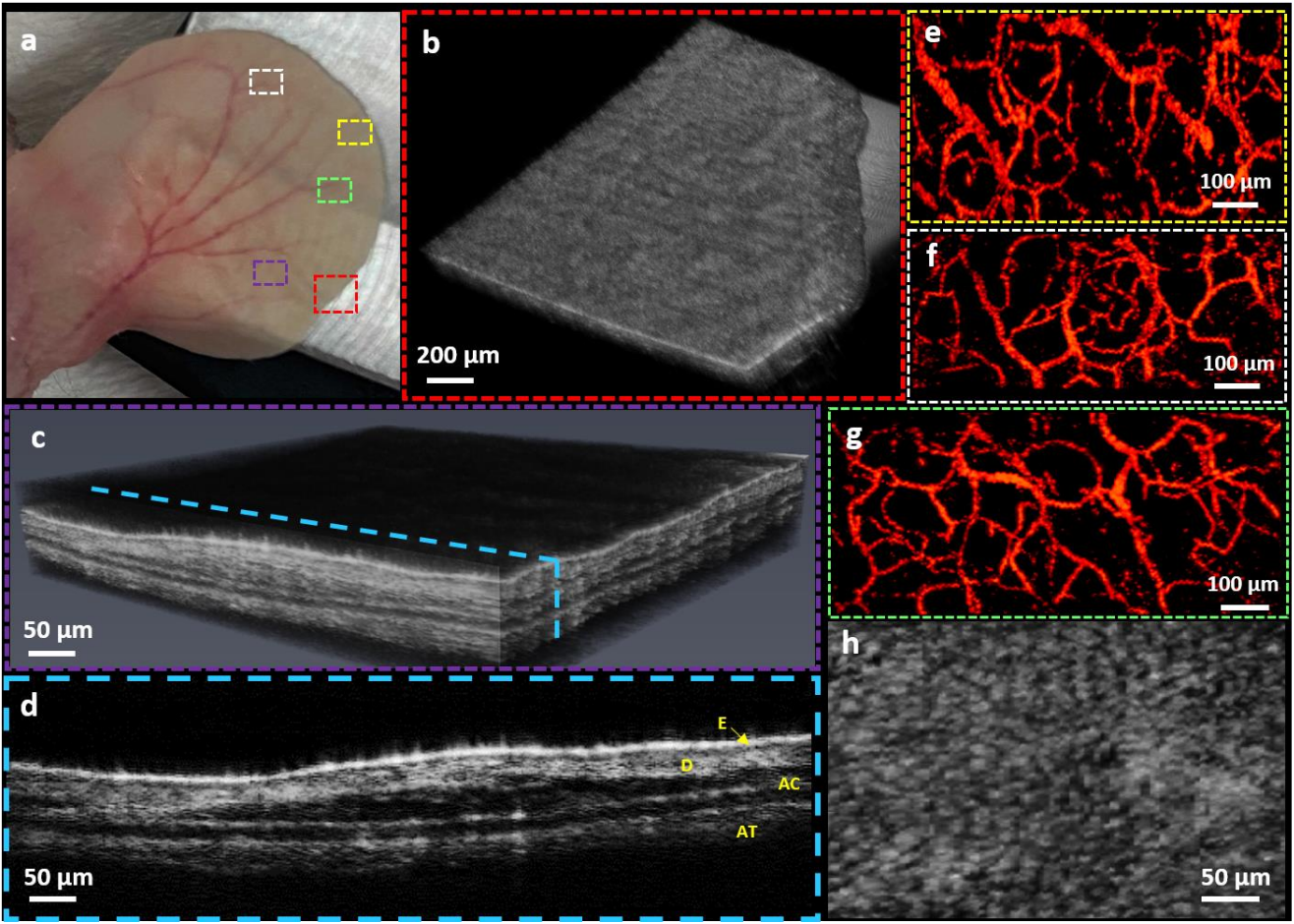

**Figure 3:** *In-vivo* imaging of mouse ear using multimodal PARS-OCT system. Corresponding white light photograph of the mouse ear, dotted rectangles indicate the PARS and OCT imaging regions (a). Volumetric rendering obtained by SS-OCT image(b). Volumetric OCT image showing different layers inside the ear tissue (c). Cross-sectional B-scan showing distinctive layers in the mouse ear tissue, E: Epidermis, D: Dermis, AC: Auricular cartilage, AT: Adipose Tissue (d). Vasculature of the ear obtained by PARS absorption mechanism (e-g). MAP image of the PARS scattering contrast showing epidermal ridges(h).

The multimodal imaging system was then applied to *in-vivo*, non-contact, structural and functional imaging of the ocular environment. First the performance of the system was tested as two standalone PARS and OCT subsystems for ocular imaging. Figure 4 shows representative images acquired using the SS-OCT system from the anterior segment of the mouse eye. Figure 4a demonstrates volumetric rendering of the entire eye, covering ~ 3.5 mm × 3.5 mm area obtained from albino mouse using a 0.26 NA microscope objective. Cross-sectional B-scans are shown in Figure 4b&c, the resolution of the SS-OCT system allows for visualizing the anterior segment structures. To bring the lower part of the anterior segment into focus and increase its signal strength, the B-scan in Figure 4b is flipped over the zero-delay line thus, iris, crystalline lens, iridocorneal angle and sclera are clearly visible. In Figure 4c, cornea epithelium layer, Bowman's membrane and stroma can be distinguished.

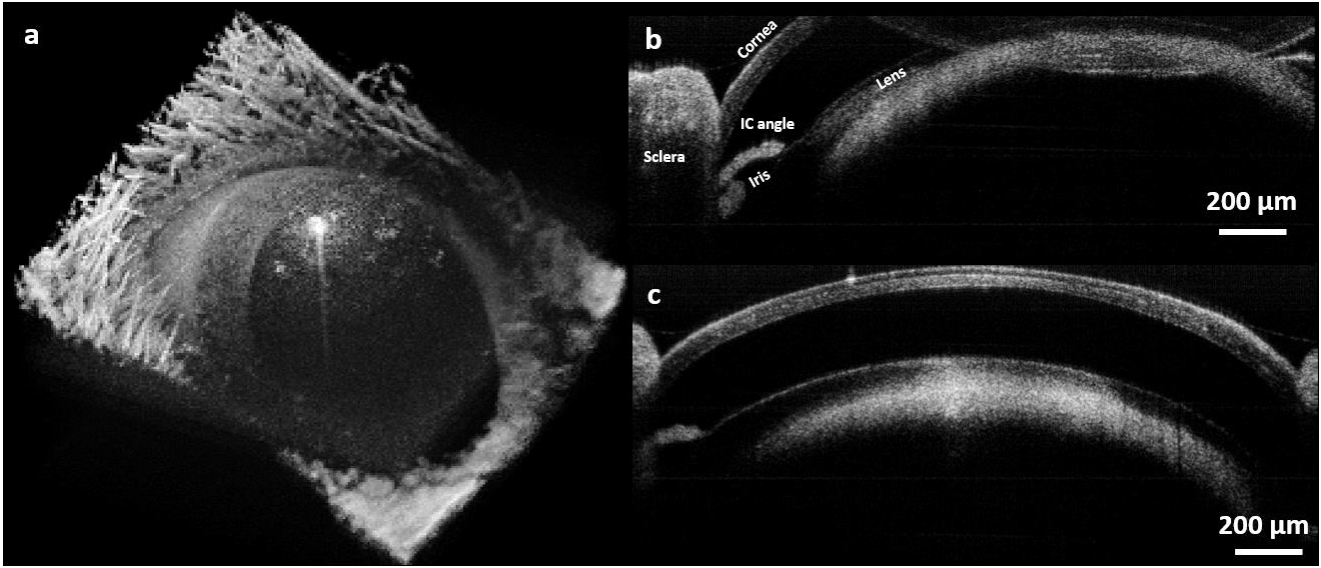

**Figure 4:** *In-vivo* imaging of ocular tissue using SS-OCT. Volumetric rendering acquired from the mouse eye (a). Cross-sectional OCT B-scan showing cornea and crystalline lens (b). Cross-sectional B-scan demonstrating cornea, sclera, crystalline lens, iris and iridocorneal angle (c).

The high-definition structural information provided by SS-OCT can be well complemented by high resolution vasculature images obtained from non-contact PARS microscopy. The potential of PARS microscopy for label-free, non-contact, *in-vivo* imaging of the iris and retinal vasculature was previously reported by our group[28]. Figure 5a shows vasculature of the eye imaged with PARS microscope. The image was acquired using a 0.26 NA objective and the field of view was measured as ~ 2.5 mm × 2.5 mm. The boxes in this figure are not highlighting the exact structure and are only used to show the approximate region imaged in Fig 5b,d&e. Figure 5b-f shown smaller field-of-views and are acquired using 20X microscope objective. The SNR of in-vivo eye images was measured ~ 35 ± 2 dB, and the in-vivo resolution of the mouse eye images was measured as 4.3 ± 0.6 μm. The mismatch between the in-vivo imaging resolution and the one measured using microbead solution can be explained by the presence of chromatic aberration and involuntary motion of the eye.

In this study, we also applied PARS microscopy for non-contact, *in-vivo* imaging of melanin which is another dominant light absorber in the ocular environment. In conventional photoacoustic imaging techniques, for *in-vivo* experiments, photoacoustic signals must propagate through various layers of the eye including the vitreous and lens which can significantly attenuate high frequency photoacoustic signals[49]. However, in PARS microscopy these pressure signals can be detected at the subsurface origin where the pressure is maximum, making PARS microscopy a great candidate for *in-vivo* ophthalmic imaging applications. Since the employed objective lens of the system has long enough working distance (~ 30 mm) compared to the diameter of mouse eyeball (~3 mm), therefore imaging the melanin content in the RPE layer was also possible using the current setup. The imaging depth of the system was measured to be ~ 3 mm corresponding for the size of the mouse eyeball. Figure 5c demonstrates representative image acquired *in-vivo* from melanin content in RPE and choroid layers through PARS microscopy. In Figure 5c, the honeycomb shaped RPE cells can be appreciated, however involuntary eye motion has distorted part of the image. Additionally, according to the Takekoshi et al. study[50], the RPE cells are more irregular in albino mice compared to pigmented ones which can partially explain the irregularity present in the images.

Capitalizing on the distinct difference in the absorption spectra of oxyhemoglobin $HbO_2$ and deoxyhemoglobin Hb, we used two excitation wavelengths (532 nm and 558 nm) to estimate the concentration of $HbO_2$ and Hb. These two wavelengths have been successfully applied for measuring $SO_2$ in previous reports[51]. The 532-nm wavelength is close to an isosbestic point, and the 558-nm wavelength is more absorptive for Hb than $HbO2$[52]. Figure 5 d&e show PARS images acquired from vasculature around the iris with 532 nm and 558 nm wavelengths, respectively. The different absorption coefficients at the two wavelengths make it possible to implement the linear $SO_2$ model. Briefly, an inverse problem is solved to estimate the relative concentrations of oxy- and deoxyhemoglobine on a per-pixel basis. The photoacoustic response is approximately linear with respect to hemoglobin concentration, and as such a regularized least squares algorithm can be used to solve for the relative concentrations. Once the relative concentrations of oxy- and deoxyhemoglobin, $[HbO_2]$ and $[Hb]$ respectively, have been obtained, oxygen saturation ($SO_2$) is estimated as $\frac{[HbO2]}{[HbO2] + [Hb]}$. Blood oxygenation levels shown in Fig 5f are pseudocolored from red to blue in an ascending order. The peripheral parts of the vessels have mainly lower oxygen saturation compared to the central regions which can be explained by oxygen uptake in the capillaries and delivery to the ocular cells[53]. The values measured in this study are an approximate estimation of blood oxygenation, and to further analyze the accuracy of PARS $SO_2$ measurement comprehensive phantom-based study using in-vitro blood will be conducted. In addition, the method will be compared with other conventional techniques such as visible light OCT and multiwavelength fundus photography.

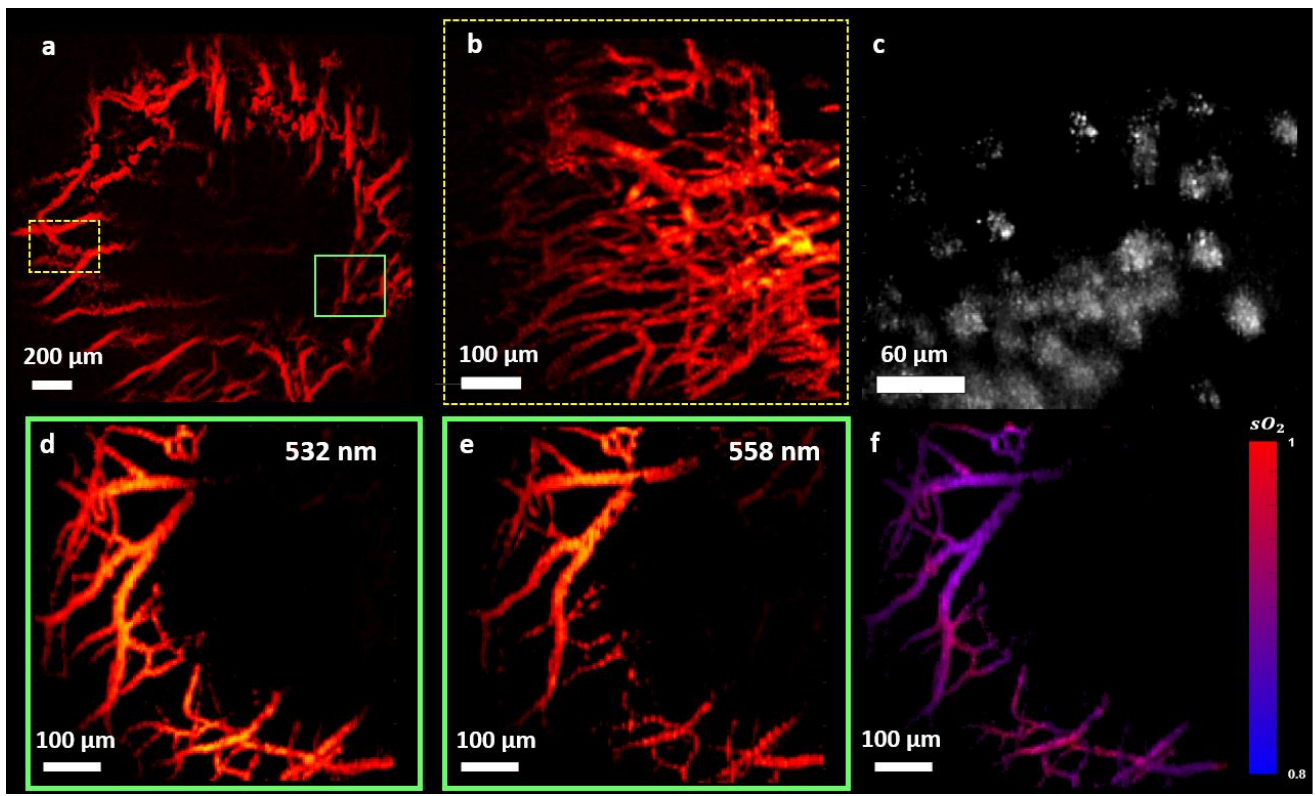

**Figure 5:** *In-vivo* imaging of ocular tissue using PARS. Vasculature of the iris imaged from ~ 2.5 mm × 2.5 mm area (a) Vasculature around the iris imaged with PARS (b) *In-vivo* images of melanin content in RPE and choroid layers (c) Images acquired using multiwavelength PARS system at 532 nm (d) 558 nm (e) and the corresponding SO$_2$ map (e).

Finally, the multimodal system is used for simultaneous imaging of the ocular tissue. Figure 6 shows representative images acquired using 0.4 NA microscope objective with the multimodal system. Figure 6a depicts OCT volume rendering of a ~ 0.7 mm ×0.7 mm area. In Fig 6b, cross-sectional B-Scan is presented, with iris tissue in focus. Yellow arrows show cross-section of the iris vasculature. Figure 6c represents e*n-face* OCT image generated from a single depth layer. The corresponding PARS image acquired simultaneously is presented in Figure 6c, yellow spline in both images represent a large vessel which appeared in both images. The overlayed image is presented in Fig 6e. Involuntary eye and head motions have distorted parts of the image in both PARS and OCT. Since the lateral resolution of the PARS subsystem is ~ 2X higher than the OCT system, smaller vasculature can be seen in the PARS image which are not clear in the OCT image. Additionally, in the current PARS setup, the detected signal is from all depth withing the detection beam depth-of-focus[24], therefore, some of the vessels in the PARS image which are not visible in the OCT *en-face* image, could be located in upper/lower layers.

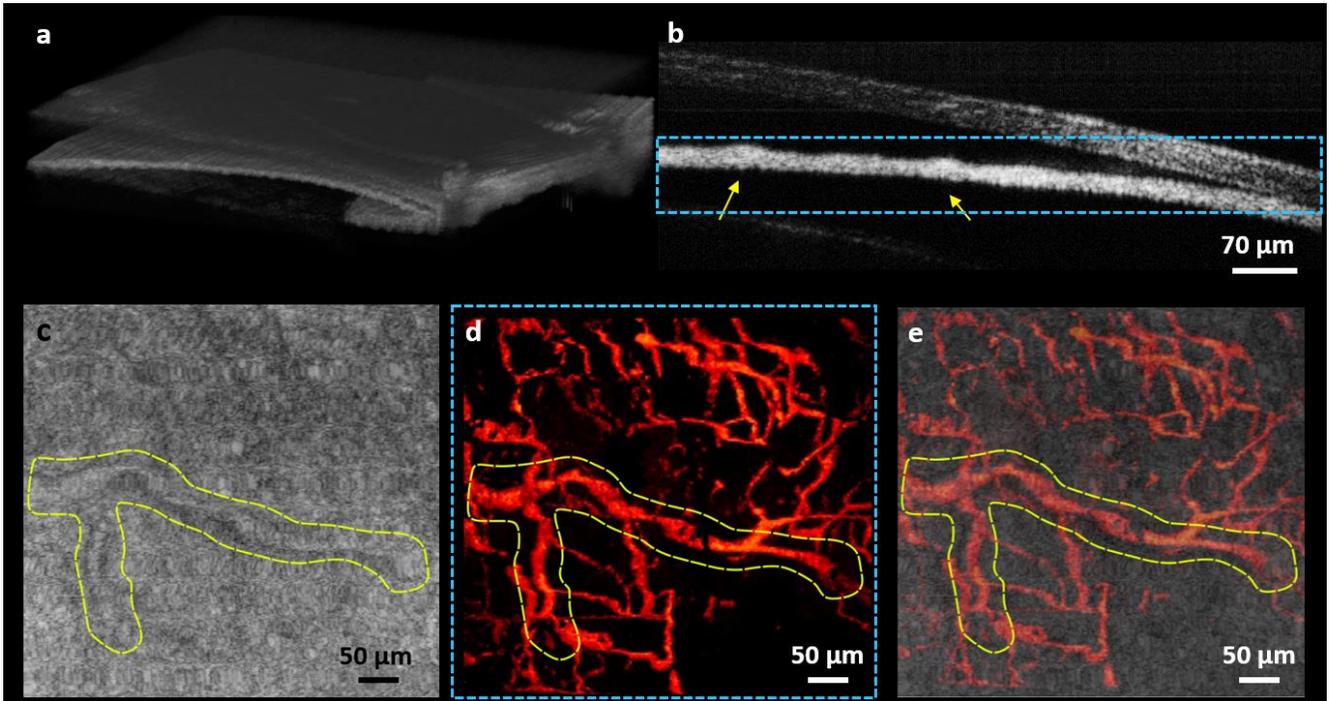

**Figure 6:** *In-vivo*, non-contact, simultaneous imaging of the ocular tissue using multimodal PARS-OCT system. Volume *OCT* rendering of the eye *(*a*).* OCT B-scan representing cross-section of the iris vasculature (yellow arrows) *(*b*)*. *en-face* OCT image *(*c*)*. vasculature of the iris imaged with the PARS subsystem *over the dashed area (*d*)*. Overlayed PARS and OCT images *(*e*)*.

There are several aspects that can be further studied in future works. We are planning to conduct a comprehensive study to evaluate the performance and sensitivity of the proposed multiwavelength PARS microscopy for accurate measurement of oxygen saturation and to compare it with other conventional techniques. Other animal models including rats and rabbits would be also employed to evaluate the performance of the system for imaging both anterior and posterior parts of the eye. In addition, to improve the field-of-view of the system required for imaging larger eyeballs we plan to employ customized scan lens. Furthermore, we are planning to quantify ocular blood flow using Doppler-OCT and PARS microscopy. These capabilities will provide us with local $SO_2$ and blood flow that will lead to oxygen metabolic rate which is a potential biomarker for major eye diseases[54].

## 4    Conclusions

In summary, for the first time, we presented *in-vivo*, non-contact, simultaneous, functional and structural ophthalmic imaging using multiwavelength PARS microscopy combined with SS-OCT. The presented multimodal system provides complementary imaging contrasts of optical absorption and optical scattering. The capability of PARS microscopy for providing both absorption and scattering imaging contrasts was presented for the first time. The non-contact imaging ability offered by both systems makes them a favorable companion for clinical ophthalmic imaging applications. The reported system is a major step toward non-invasive, simultaneous, and accurate measurement of functional details in the ophthalmic tissue and can assist ophthalmologists with the diagnostics and treatment of major eye diseases.

**Acknowledgements:**

The authors would like to thank Jean Flanagan for assistance with animal-related procedures. The authors would also like to thank Kevan Bell, Benjamin Ecclestone, and Marian Boktor for their support and help. The authors acknowledge funding from New Frontiers in Research Fund – Exploration (NFRFE-2019-01012); Natural Sciences and Engineering Research Council of Canada (DGECR-2019-00143, RGPIN2019-06134); Canada Foundation for Innovation (JELF #38000); Mitacs (IT13594); Centre for Bioengineering and Biotechnology (CBB Seed fund); University of Waterloo and illumiSonics (SRA #083181).

**Additional Information**

*Conflict of interest:*

Author P. Haji Reza has financial interests in illumiSonics Inc. IllumiSonics partially supported this work. All other authors have no competing/conflict of interest.

**References**
1. Liu, W. & Zhang, H. F. Photoacoustic imaging of the eye: A mini review. *Photoacoustics* **4**, 112–123 (2016).
2. Hammer, M. *et al.* Diabetic patients with retinopathy show increased retinal venous oxygen saturation. *Graefes Arch Clin Exp Ophthalmol* **247**, 1025–1030 (2009).
3. Khoobehi, B., Firn, K., Thompson, H., Reinoso, M. & Beach, J. Retinal Arterial and Venous Oxygen Saturation Is Altered in Diabetic Patients. *Invest. Ophthalmol. Vis. Sci.* **54**, 7103–7106 (2013).
4. Vandewalle, E. *et al.* Oximetry in glaucoma: correlation of metabolic change with structural and functional damage. *Acta Ophthalmologica* **92**, 105–110 (2014).
5. Olafsdottir, O. B., Hardarson, S. H., Gottfredsdottir, M. S., Harris, A. & Stefánsson, E. Retinal Oximetry in Primary Open-Angle Glaucoma. *Invest. Ophthalmol. Vis. Sci.* **52**, 6409–6413 (2011).
6. Hardarson, S. H. & Stefánsson, E. Oxygen Saturation in Central Retinal Vein Occlusion. *American Journal of Ophthalmology* **150**, 871–875 (2010).
7. Eliasdottir, T., Bragason, D., Hardarson, S. & Stefánsson, E. Retinal Oxygen Saturation is Affected in Central Retinal Vein Occlusion. *Invest. Ophthalmol. Vis. Sci.* **54**, 46–46 (2013).
8. HYMAN, L. G., LILIENFELD, A. M., FERRIS, F. L., III & FINE, S. L. SENILE MACULAR DEGENERATION: A CASE-CONTROL STUDY. *American Journal of Epidemiology* **118**, 213–227 (1983).

9. Landrum, J., Bone, R. & Kilburn, M. The macular pigment: A possible role in protection from age-related macular degeneration. **38**, 537–538 (1997).
10. Yi, J. *et al.* Visible light optical coherence tomography measures retinal oxygen metabolic response to systemic oxygenation. *Light: Science & Applications* **4**, e334–e334 (2015).
11. Lau, J. C. M. & Linsenmeier, R. A. Increased intraretinal PO2 in short-term diabetic rats. *Diabetes* **63**, 4338–4342 (2014).
12. Chen, S. *et al.* Retinal oximetry in humans using visible-light optical coherence tomography [Invited]. *Biomed Opt Express* **8**, 1415–1429 (2017).
13. Finikova, O. S. *et al.* Oxygen microscopy by two-photon-excited phosphorescence. *Chemphyschem* **9**, 1673–1679 (2008).
14. Ito, Y. & Berkowitz, B. A. MR studies of retinal oxygenation. *Vision Res* **41**, 1307–1311 (2001).
15. Pittman, R. N. *Measurement of Oxygen*. *Regulation of Tissue Oxygenation* (Morgan & Claypool Life Sciences, 2011).
16. Schweitzer, D., Thamm, E., Hammer, M. & Kraft, J. A new method for the measurement of oxygen saturation at the human ocular fundus. *International Ophthalmology* **23**, 347–353 (2001).
17. Human macular pigment assessed by imaging fundus reflectometry. *Vision Research* **29**, 663–674 (1989).
18. Hosseinaee, Z., Le, M., Bell, K. & Haji Reza, P. Towards non-contact photoacoustic imaging [Review]. *Photoacoustics* 100207 (2020) doi:10.1016/j.pacs.2020.100207.
19. Yao, J. & Wang, L. V. Sensitivity of photoacoustic microscopy. *Photoacoustics* **2**, 87–101 (2014).
20. Shu, X., Li, H., Dong, B., Sun, C. & Zhang, H. F. Quantifying melanin concentration in retinal pigment epithelium using broadband photoacoustic microscopy. *Biomed. Opt. Express, BOE* **8**, 2851–2865 (2017).
21. Chen, S., Yi, J. & Zhang, H. F. Measuring oxygen saturation in retinal and choroidal circulations in rats using visible light optical coherence tomography angiography. *Biomed Opt Express* **6**, 2840–2853 (2015).
22. Song, W. *et al.* A combined method to quantify the retinal metabolic rate of oxygen using photoacoustic ophthalmoscopy and optical coherence tomography. *Sci Rep* **4**, 6525 (2015).
23. Link, D. *et al.* Novel non-contact retina camera for the rat and its application to dynamic retinal vessel analysis. *Biomed Opt Express* **2**, 3094–3108 (2011).
24. Haji Reza, P., Shi, W., Bell, K., Paproski, R. J. & Zemp, R. J. Non-interferometric photoacoustic remote sensing microscopy. *Light: Science & Applications* **6**, e16278–e16278 (2017).
25. Ecclestone, B. R. *et al.* Improving maximal safe brain tumor resection with photoacoustic remote sensing microscopy. *Scientific Reports* **10**, 17211 (2020).
26. Ecclestone, B. R. *et al.* Towards virtual biopsies of gastrointestinal tissues using photoacoustic remote sensing microscopy. *Quantitative Imaging in Medicine and Surgery* **11**, 1070077–1071077 (2021).
27. Haji Reza, P., Bell, K., Shi, W., Shapiro, J. & Zemp, R. J. Deep non-contact photoacoustic initial pressure imaging. *Optica* **5**, 814–820 (2018).
28. Hosseinaee, Z. *et al.* Label-free, non-contact, in vivo ophthalmic imaging using photoacoustic remote sensing microscopy. *Opt. Lett., OL* **45**, 6254–6257 (2020).
29. de Boer, J. F., Leitgeb, R. & Wojtkowski, M. Twenty-five years of optical coherence tomography: the paradigm shift in sensitivity and speed provided by Fourier domain OCT [Invited]. *Biomed. Opt. Express* **8**, 3248 (2017).
30. Hosseinaee, Z., Tummon Simmons, J. A. & Haji Reza, P. Dual-Modal Photoacoustic Imaging and Optical Coherence Tomography [Review]. *Front. Phys.* **8**, (2021).
31. OSA | Multimodal imaging with spectral-domain optical coherence tomography and photoacoustic remote sensing microscopy. https://www-osapublishing-org.proxy.lib.uwaterloo.ca/ol/abstract.cfm?uri=ol-45-17-4859.
32. Wang, X. *et al.* Noninvasive photoacoustic angiography of animal brains in vivo with near-infrared light and an optical contrast agent. *Opt. Lett., OL* **29**, 730–732 (2004).
33. Park, S. M. *et al.* Quickly Alternating Green and Red Laser Source for Real-time Multispectral Photoacoustic Microscopy. *Photoacoustics* **20**, 100204 (2020).
34. Haji Reza, P., Forbrich, A. & Zemp, R. In-Vivo functional optical-resolution photoacoustic microscopy with stimulated Raman scattering fiber-laser source. *Biomed. Opt. Express* **5**, 539 (2014).
35. Haji Reza, P., Forbrich, A. & Zemp, R. J. Multifocus optical-resolution photoacoustic microscopy using stimulated Raman scattering and chromatic aberration. *Opt. Lett., OL* **38**, 2711–2713 (2013).

36. Abu-Sardanah, S. O., Subramaniam, C., Abbasi, S. & Haji Reza, P. A comprehensive characterization of a stimulated Raman scattering fiber-laser source for multi-wavelength dependent photoacoustic microscopy techniques (Conference Presentation). in *Photons Plus Ultrasound: Imaging and Sensing 2020* vol. 11240 1124024 (International Society for Optics and Photonics, 2020).
37. Liu, C., Chen, J., Zhang, Y., Zhu, J. & Wang, L. Five-wavelength optical-resolution photoacoustic microscopy of blood and lymphatic vessels. *AP* **3**, 016002 (2021).
38. Ahsen, O. O. *et al.* Swept source optical coherence microscopy using a 1310 nm VCSEL light source. *Opt. Express, OE* **21**, 18021–18033 (2013).
39. Chen, L. & Xu, J. OPTIMAL DELAUNAY TRIANGULATIONS. *Journal of Computational Mathematics* **22**, 299–308 (2004).
40. Cense, B. *et al.* Ultrahigh-resolution high-speed retinal imaging using spectral-domain optical coherence tomography. *Opt. Express, OE* **12**, 2435–2447 (2004).
41. Schneider, C. A., Rasband, W. S. & Eliceiri, K. W. NIH Image to ImageJ: 25 years of image analysis. *Nature Methods* **9**, 671–675 (2012).
42. ANSI Z136.1-2014: Safe Use of Lasers - ANSI Blog. *The ANSI Blog* https://blog.ansi.org/2015/07/ansi-z1361-2014-safe-use-of-lasers/ (2015).
43. Delori, F. C., Webb, R. H. & Sliney, D. H. Maximum permissible exposures for ocular safety (ANSI 2000), with emphasis on ophthalmic devices. *J. Opt. Soc. Am. A, JOSAA* **24**, 1250–1265 (2007).
44. Jeon, S. *et al.* In Vivo Photoacoustic Imaging of Anterior Ocular Vasculature: A Random Sample Consensus Approach. *Sci Rep* **7**, 4318 (2017).
45. Hariri, S. *et al.* Limiting factors to the OCT axial resolution for in-vivo imaging of human and rodent retina in the 1060nm wavelength range. *Opt. Express* **17**, 24304 (2009).
46. Eriksson, E., Boykin, J. V. & Pittman, R. N. Method for in vivo microscopy of the cutaneous microcirculation of the hairless mouse ear. *Microvasc Res* **19**, 374–379 (1980).
47. Haindl, R. *et al.* Functional optical coherence tomography and photoacoustic microscopy imaging for zebrafish larvae. *Biomed. Opt. Express, BOE* **11**, 2137–2151 (2020).
48. Li, L., Maslov, K., Ku, G. & Wang, L. V. Three-dimensional combined photoacoustic and optical coherence microscopy for in vivo microcirculation studies. *Optics express* **17**, 16450–16455 (2009).
49. Ye, S. G., Harasiewicz, K. A., Pavlin, C. J. & Foster, F. S. Ultrasound characterization of normal ocular tissue in the frequency range from 50 MHz to 100 MHz. *IEEE Transactions on Ultrasonics, Ferroelectrics, and Frequency Control* **42**, 8–14 (1995).
50. L, I.-T. *et al.* Retinal pigment epithelial integrity is compromised in the developing albino mouse retina. *J Comp Neurol* **524**, 3696–3716 (2016).
51. Cao, R. *et al.* Photoacoustic microscopy reveals the hemodynamic basis of sphingosine 1-phosphate-induced neuroprotection against ischemic stroke. *Theranostics* **8**, 6111–6120 (2018).
52. Liu, C., Liang, Y. & Wang, L. Optical-resolution photoacoustic microscopy of oxygen saturation with nonlinear compensation. *Biomed Opt Express* **10**, 3061–3069 (2019).
53. Hogeboom van Buggenum, I. M., van der Heijde, G. L., Tangelder, G. J. & Reichert-Thoen, J. W. Ocular oxygen measurement. *Br J Ophthalmol* **80**, 567–573 (1996).
54. Wang, L. V. Prospects of photoacoustic tomography. *Med Phys* **35**, 5758–5767 (2008).

**Author Contributions:**

Z.H. and N.A. constructed the PARS-OCT system. Z.H. wrote the manuscript and prepared the figures. N.P. developed and implemented the signal unmixing method. L.M and L.K. assist with animal preparation and conducting imaging experiments. P.H.R. conceived the project and acted as the primary investigator. All the authors reviewed the manuscript.